\documentclass{aa}  
\usepackage{natbib}
\usepackage{epstopdf}

\usepackage{epsfig}
\usepackage{amsmath,amssymb,amstext}
\usepackage[english]{babel}
\usepackage[utf8]{inputenc}
\usepackage[flushleft]{threeparttable}
\usepackage{float}
\usepackage{placeins}
\usepackage{graphicx}
\usepackage{hyperref}
\usepackage{url}
\usepackage{txfonts}
\usepackage{multirow}
\usepackage{microtype}
\usepackage{capt-of}

\newcommand{\D}{\mathrm{d}}

\begin{document}

   \title{Disentangling the ISM phases of the dwarf galaxy NGC 4214 using [\ion{C}{ii}] SOFIA/GREAT observations\thanks{Reduced SOFIA/GREAT data is available
at the CDS via anonymous ftp to cdsarc.u-strasbg.fr (130.79.128.5)
or via \url{http://cdsweb.u-strasbg.fr/cgi-bin/qcat?J/A+A/}}}
\titlerunning{Disentangling the ISM phases of the dwarf galaxy NGC 4214}

   \author{K. Fahrion \inst{1}
           \and
           D. Cormier\inst{1} 
            \and
           F. Bigiel\inst{1}
           \and
           S. Hony\inst{1}
           \and
           N. P. Abel\inst{2}
           \and
           P. Cigan\inst{3}
           \and
           T. Csengeri\inst{4}
           \and
           U. U. Graf\inst{5}
           \and
           V. Lebouteiller\inst{6}
           \and 
           S. C. Madden\inst{6}
           \and
           R. Wu\inst{7}           
           \and
           L. Young\inst{8}
          }

   \institute{Institut f\"ur Theoretische Astrophysik, Zentrum f\"ur Astronomie der Universit\"at Heidelberg,
              Albert-\"Uberle-Stra\ss e 2, 69120 Heidelberg, Germany \\\email{fahrion@stud.uni-heidelberg.de}
     \and  
University of Cincinnati, Clermont College, MCGP Department, Batavia, OH 45103, USA   
     \and
     School of Physics and Astronomy, Cardiff University, Cardiff CF24 3AA, UK
     \and
     	Max-Planck-Institut f\"ur Radioastronomie, Auf dem H\"ugel 69,
53121 Bonn, Germany
	\and
	I. Physikalisches Institut der Universität zu K\"oln, Z\"ulpicher
Straße 77, 50937 K\"oln, Germany
\and
     Laboratoire AIM, CEA Saclay, Orme des Merisiers, Bat 709, 91 191 Gif-sur-Yvette, France
	\and
	International Research Fellow of the Japan Society for the Promotion of Science (JSPS), Department of Astronomy, the University of Tokyo, Bunkyo-ku, 113-0033 Tokyo, Japan
	\and
	Physics Department, New Mexico Tech, 801 Leroy Place, Socorro NM 87801 USA
	}

   \date{\today}

  \abstract
{The [\ion{C}{ii}] 158 $\mu$m fine structure line is one of the dominant cooling lines in the interstellar medium (ISM) and is an important tracer of star formation. Recent velocity-resolved  studies with \textit{Herschel}/HIFI and SOFIA/GREAT showed that the [\ion{C}{ii}] line can constrain the properties of the ISM phases in star-forming regions. The [\ion{C}{ii}] line as a tracer of star formation is particularly important in low-metallicity environments where CO emission is weak because of the presence of large amounts of CO-dark gas.}
{The nearby irregular dwarf galaxy NGC 4214 offers an excellent opportunity to study an actively star-forming ISM at low metallicity. We analyzed the spectrally resolved [\ion{C}{ii}] line profiles in three distinct regions at different evolutionary stages of NGC 4214 with respect to ancillary \ion{H}{i} and CO data in order to study the origin of the [\ion{C}{ii}] line.}
{We used SOFIA/GREAT [\ion{C}{ii}] 158 $\mu$m observations, \ion{H}{i} data from THINGS, and CO(2 $\rightarrow$ 1) data from HERACLES to decompose the spectrally resolved [\ion{C}{ii}] line profiles into components associated with neutral atomic and molecular gas. We use this decomposition to infer gas masses traced by [\ion{C}{ii}] under different ISM conditions.}
{Averaged over all regions, we associate about 46\% of the [\ion{C}{ii}] emission with the \ion{H}{i} emission. However, we can assign only $\sim$ 9 \% of the total [\ion{C}{ii}] emission to the cold neutral medium (CNM). We found that about 79\% of the total molecular hydrogen mass is not traced by CO emission.} 
{On average, the fraction of CO-dark gas dominates the molecular gas mass budget. The fraction seems to depend on the evolutionary stage of the regions: it is highest in the region covering a super star cluster in NGC 4214, while it is lower in a more compact, more metal-rich region.}
   \maketitle

\section{Introduction}
\begin{figure*}
\centering
\includegraphics[width=0.43\textwidth]{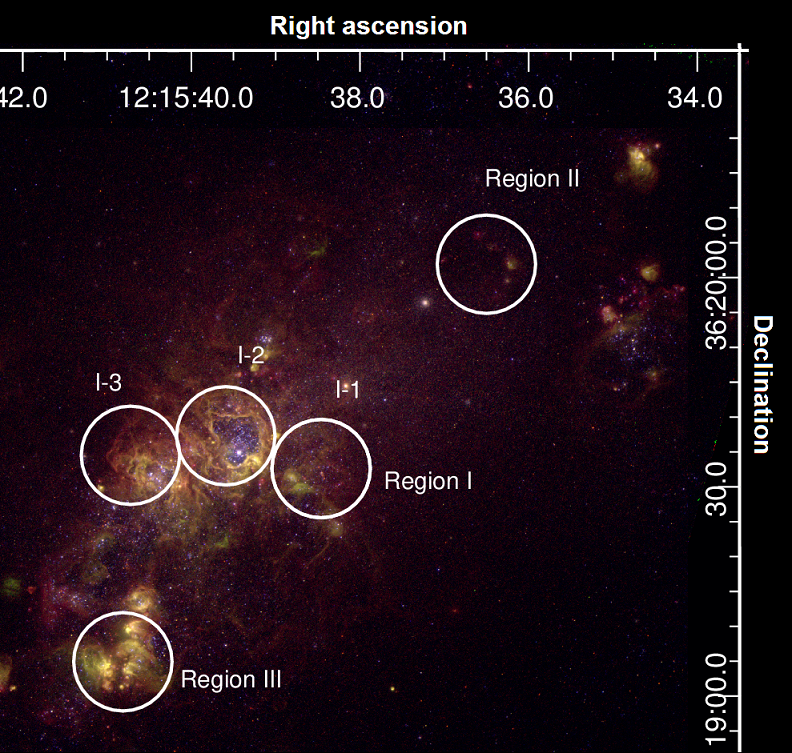}
\hspace{1cm}
\includegraphics[width=0.43\textwidth]{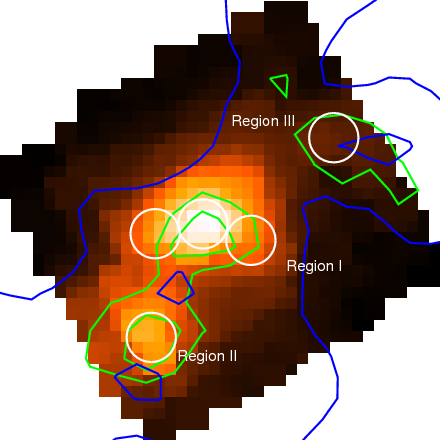}
\caption{Left: Three-color image of the central part of NGC 4214 and pointing positions for SOFIA/GREAT observations. HST WFC3 filters F336W (blue), F502 (green), and F657 (red) were used. Credit: NASA/Hubble. Right: Integrated [\ion{C}{ii}] intensity map observed with \textit{Herschel}/PACS \citep{Cormier2010}. \ion{H}{i} \citep{ThingsPaper} contours overlayed in blue, CO(2 $\rightarrow$ 1) \citep{HeraclesPaper} contours in green, and SOFIA/GREAT pointings as white circles with a diameter corresponding to 14.1 \arcsec ($\sim$ 200 pc). For CO(2 $\rightarrow$ 1), the contour levels are at 1.2 K km s$^{-1}$ and 2.4 K km s$^{-1}$. For the extended \ion{H}{i} the levels are at 0.23 Jy beam$^{-1}$ km s$^{-1}$ and 0.27 Jy beam$^{-1}$ km s$^{-1}$. The map covers a 1.6\arcmin $\times$ 1.6\arcmin field of view ($\sim$ 1.3 kpc $\times$ 1.3 kpc).}
\label{fig:HubbleNGC4214}
\end{figure*}

Star formation proceeds through the contraction and collapse of cold, dense molecular clouds. To study the first stages of star formation, observations of these molecular clouds and the knowledge of their conditions are of great interest. However, direct observations of the most abundant molecule in the cold molecular clouds, H$_2$, is not possible because the H$_2$ molecule has no dipole moment and can therefore not be observed in cold and dense environments. Instead, carbon monoxide is traditionally used to study the molecular phase of the interstellar medium (ISM) as CO forms under the conditions prevalent in cold molecular clouds. The lowest ground rotational transition J=1 $\rightarrow$ 0 creates the most prominent line at 2.6 mm (115.27 GHz), but various other transition lines can be observed as well. The reservoir of molecular hydrogen from which new stars form is usually estimated via the CO-to-H$_{2}$ conversion factor $\alpha_{\mathrm{CO}}$ (or $X_{\mathrm{CO}}$).
Nonetheless, the CO-to-H$_{2}$ conversion factor is a controversial quantity as it seems to vary with metallicity \citep{Bolatto2013,Poglitsch1995,Cormier2014, Narayanan2011,GloverMaclow2011}. Furthermore, evidence for hidden molecular hydrogen ("CO-Dark Molecular Gas" \citealt{Wolfire2010}) has beed found not only in the Milky Way \citep{GOTCCO, Planck2011, Grenier2005}, but also in other galaxies \citep{Israel1997,Madden1997,Requena-Torres2016}.

These recent studies raise the importance of other tracers that might give insight into the conditions of the molecular content. One of the most promising tracers is the [\ion{C}{ii}] fine-structure line at 158 $\mu$m emitted by singly ionized carbon. [\ion{C}{ii}] is usually the strongest far-infrared line, representing up to 1\% of the continuum emission in the FIR regime (e.g., \citealt{Stacey1991}, \citealt{Genzel2000}, \citealt{Cormier2015}). Being one of the dominant cooling lines in photon-dominated regions (PDRs, \citealt{HollenbachTielens}) it is often used to study their conditions. 

However, because ionized carbon dominates the carbon abundance under most conditions found in the ISM and the [\ion{C}{ii}] line can be excited by collisions with electrons, atomic, or molecular hydrogen, this line not only traces PDRs, but it can also be found in other phases of the ISM (e.g., \citealt{Goldsmith2012}). Therefore, the origin of the [\ion{C}{ii}] line has to be studied in detail in order to use it as a tracer of star-forming regions. Disentangling the different contributions from molecular, ionized, or atomic phases can be done by comparing the spectrally or spatially resolved [\ion{C}{ii}] emission with tracers of other ISM phases. As high spatial and spectral resolution is needed, this can only be done in nearby objects. So far, integrated line intensities have mostly been used to study star formation in external galaxies, but velocity-resolved observations of the Milky Way and nearby galaxies have become available with the HIFI instrument on $Herschel$ and with the GREAT instrument on SOFIA. Constraining the amount of CO-dark molecular gas using [\ion{C}{ii}] is of
particular interest for low-metallicity dwarf galaxies. These galaxies can have very weak CO emission \citep{Schruba2012, Cormier2014}.   
At the same time, dwarf galaxies can produce stars at rates that are normally found in starburst galaxies (e.g., \citealt{Gallagher1984}, \citealt{Hunter1989}). These two observations combined imply that the CO-dark gas may
outweigh the CO-bright gas by a large factor. Measuring the CO-dark gas fraction in low-metallicity dwarf galaxies is therefore of key importance for understanding their behavior.

In this paper, we take advantage of the new generation of velocity-resolved [\ion{C}{ii}] observations from SOFIA/GREAT to disentangle the origin of [\ion{C}{ii}] emission in NGC 4214. NGC 4214 is a nearby irregular dwarf galaxy of Magellanic type at a distance of 2.9 Mpc \citep{MaApell2002, Dalcanton2009} with low metallicity (log(O/H) + 12 = 8.2, \citealt{Kobulnicky1996}). 
Owing to its proximity, there is a wealth of ancillary data for NGC 4214. CO(1 $\rightarrow$ 0) observations showed that NGC 4214 exhibits three distinct star-forming regions in different evolutionary stages \citep{Walter2001}. 
The left panel in Figure \ref{fig:HubbleNGC4214} shows a three-color image of the central part of NGC 4214 with the pointing positions of our SOFIA/GREAT observations of the star-forming regions overlayed as white circles. In the following, we use the labeling of the regions as shown in Figure \ref{fig:HubbleNGC4214}. 
The central region (region I) is the most evolved and is also the largest region, containing a super star cluster and several star clusters \citep{MaizApellaniz2001}. Bright O stars as well as Wolf-Rayet stars and a H$\alpha$ cavity  are found in this region \citep{MacKenty2000}. Using a UV-optical survey of NGC 4214, \cite{Ubeda2007Results} showed that region I is about $\sim 3 - 4$ Myr old. 
In contrast, the southeastern region, labeled as region II, is only $\sim$ 2 Myr of age. This region is more compact and contains concentrated 
gas reservoirs \citep{MacKenty2000} as well as OB stars and strong H$\alpha$ emission, which is created in bright H$\alpha$ knots
\citep{Ubeda2007Methods, MaApell1998}. Using H$\alpha$ images and \ion{H}{i} velocity data obtained by \cite{McIntyre1997}, \cite{Thurow2005} found systematic differences in the kinematics of the ionized gas in region I and II.
In the image, the southeastern region is labeled as region III. No star formation is recognizable in this region, but CO emission was found by \cite{Walter2001}. Since this region is the least evolved,
\cite{Walter2001} suggest that region III will form stars in the future. 
With these three distinct regions, NGC 4214 offers the possibility of studying the evolution of star formation at low metallicities and in different environments.

We present observations of [\ion{C}{ii}] emission in three regions of NGC 4214 using the GREAT instrument on board the SOFIA telescope. Our main objective is to extend the few studies of CO-dark gas in galaxies by decomposing the spectrally resolved [\ion{C}{ii}] emission of NGC 4214 into fractions associated with the atomic and the molecular hydrogen gas content. 
In particular, our new observations allow us to determine which
fraction of the [\ion{C}{ii}] emission can be associated spectrally with
the CO-emitting gas and, in turn, which fraction is too far removed
in velocity-space to be physically linked directly to the cold dense
gas detected in CO.
In Section \ref{sec:Data} we describe the data, in Section \ref{sec:Decomp} we describe our decomposition method with its results, and in Section \ref{sec:AssociatedGas} we present the associated gas fractions and masses. Our analysis is summarized and discussed in Section \ref{sec:Discussion}.

\section{Data}
\label{sec:Data}
\subsection{\texorpdfstring{[\ion{C}{ii}] data}{[CII] data}}

\subsubsection{SOFIA data}
The [\ion{C}{ii}] emission at 1900.54 GHz (158 $\mu$m) was observed in NGC 4214 using the L2 band of the GREAT instrument \citep{Heyminck2012} on board the airborne Stratospheric Observatory for Infrared Astronomy (SOFIA; \citealt{SOFIA}). Observations were performed as part of observation cycle 2 in May 2014 and January 2015 with a half-power beam width of 14.1\,\arcsec ($\sim$ 200 pc) and 1.16 $\mathrm{km \, s}^{-1}$ velocity resolution. 
Five positions in NGC 4214 were observed in the three main star-forming regions (Figure \ref{fig:HubbleNGC4214}; the coordinates can be found in Table \ref{tab:Fluxes}). SOFIA has a pointing accuracy of $\sim$ 0.5\arcsec.
The data were processed with the eXtended bandwidth Fast Fourier Transform Spectrometer (XFFTS) and the standard GREAT calibrator \citep{Guan2012}. The data were calibrated to antenna temperature scale ($\eta_f = 0.97$) and to the main beam temperature scale ($\eta_{mb} = 0.65 - 0.69$ for L2). We converted the data from main beam temperature scale to Jansky using a conversion factor of 585 Jy/K. 

Before averaging the spectra, the baseline was removed using a polynomial of first order except in a few spectra from May 2014 where third-order polynomials were used. Data reduction was done using the CLASS software\footnote{\url{http://www.iram.fr/IRAMFR/GILDAS}}. For further analysis we rebinned the data to 2.6 $\mathrm{km \, s}^{-1}$ resolution.
The fluxes reported in Table \ref{tab:Fluxes} were calculated by direct integration of the rebinned spectra between 270 and 330 $\mathrm{km \, s}^{-1}$, except for region III where we used channels between 240 and 330 $\mathrm{km \, s}^{-1}$. These ranges were chosen based on the line widths also observed in CO and \ion{H}{i}.

\subsubsection{Herschel/PACS}
\label{subsec:PACS}
NGC 4214 was also observed in the [\ion{C}{ii}] line with the PACS spectrometer \citep{PACS} on the \textit{Herschel} Space Observatory \citep{Herschel}. The data were originally presented in \cite{Cormier2010} and reprocessed with the PACS spectrometer pipeline of the \textit{Herschel} interactive processing environment (HIPE) v14.2. The PACS map covers a 1.6\arcmin $\times$ 1.6\arcmin\, field of view with a spatial resolution of 12\arcsec.  Figure \ref{fig:HubbleNGC4214} shows the integrated intensity map observed with \textit{Herschel}/PACS in the right panel. We  compare the fluxes of GREAT and PACS for the regions observed with both instruments.
To this end, the PACS map was convolved with a Gaussian kernel to the angular resolution of SOFIA/GREAT data and aperture photometry was applied to compute the fluxes shown in Table \ref{tab:Fluxes}. For both [\ion{C}{ii}] datasets, the emission is strongest in pointing I-2 in the central region of NGC 4214 where the super star cluster is located and weakest in region III. The \textit{Herschel}/PACS and SOFIA/GREAT fluxes are in relatively good agreement (within 40\%) in regions I and II. In region III, the GREAT flux is about two times higher than the PACS flux, but [\ion{C}{ii}] is not detected at high significance in this region.

\subsection{\texorpdfstring{CO and \ion{H}{i} data}{CO and HI data}}
\label{subsec:COandHI}
NGC 4214 was observed in the CO(1 $\rightarrow$ 0) transition line  with the Caltech OVRO millimeter interferometer by \cite{Walter2001}. The data have a velocity resolution of 1.30 $\mathrm{km \, s}^{-1}$ and the half-power beam width is 6.4$'' \, \times 5.7''$. 

NGC 4214 was also included in the HERA CO-Line Extragalactic Survey (HERACLES; \citealt{HeraclesPaper}). The CO(2 $\rightarrow$ 1) line at 230.54 GHz (1.3 mm) was observed with a half-power beam width of 13\arcsec and 2.6 $\mathrm{km \, s}^{-1}$ velocity resolution.
Although the spectral resolution is higher for the OVRO CO(1 $\rightarrow$ 0) data, we used CO(2 $\rightarrow$ 1) to distinguish the origin of [\ion{C}{ii}] emission because the velocity range covered by the OVRO data is narrower, ranging only from 262 to 343 $\mathrm{km \, s}^{-1}$, while the \ion{H}{i} emission is more extended than this. This can be seen in Figure \ref{fig:spectra}. A second disadvantage of using the OVRO interferometry data is that we might miss extended CO emission. However, in the low-metallicity environment of NGC 4214, we do not expect a significant amount of extended emission. 

Nonetheless, we want to investigate the possibility of missing CO in the OVRO observations and thus calculate the flux ratios of CO(1 $\rightarrow$ 0) and CO(2 $\rightarrow$ 1).   
Table \ref{tab:Fluxes} provides the flux ratios (in K km s$^{-1}$) for both OVRO and HERACLES datasets in each region. To calculate this quantity, we convolved the OVRO CO(1 $\rightarrow$ 0) data to the angular resolution of the HERACLES observation and integrated by summation over the channels. Especially in region I, the values are higher than the typical ratios of 0.8 \citep{Braine1993}, but overall they agree within the uncertainties. The values seem reasonable, but we note that the OVRO data may be subject
to missing flux (zero-spacing) so that the ratios formally represent an upper limit of the true ratio. However, given the
strong radiation fields from ongoing star formation - in particular in regions I and II - and the low-metallicity, we do not expect a diffuse CO component.

We used \ion{H}{i} data from The HI Nearby Galaxy Survey (THINGS; \citealt{ThingsPaper}) which performed observations of the atomic hydrogen content in nearby galaxies using the 21 cm (1.4204 GHz) line of atomic hydrogen with a half-power beam width of 14.69\arcsec $\times$ 13.87\arcsec and 1.29 $\mathrm{km \, s}^{-1}$ spectral resolution.

Table \ref{tab:Fluxes} lists the fluxes for \ion{H}{i}, CO(2 $\rightarrow$ 1), and CO(1 $\rightarrow$ 0) in all regions. The fluxes were derived after convolving the CO data to the angular resolution of SOFIA/GREAT. For the CO lines, we chose channels between 270 and 330 $\mathrm{km \, s}^{-1}$ for integration and for \ion{H}{i} we take the channels between 250 and 350 $\mathrm{km \, s}^{-1}$. In region III, however, we set the integration range to 240 - 300 $\mathrm{km \, s}^{-1}$ for both CO and \ion{H}{i} spectra. The \ion{H}{i} fluxes are almost constant over the regions, while the CO fluxes have their maximum in region II.

\subsection{Description of the spectra}
The spectra of [\ion{C}{ii}], \ion{H}{i}, CO(2 $\rightarrow$ 1), and CO(1 $\rightarrow$ 0) in the five regions observed by SOFIA/GREAT are shown in Figure \ref{fig:spectra}. 

By fitting Gaussians, we find that the [\ion{C}{ii}] emission lines have widths of $\sim$ 20 $\mathrm{km \, s}^{-1}$ and are centered at about 300 $\mathrm{km \, s}^{-1}$, except in region III where the data show weak [\ion{C}{ii}] emission centered at lower velocities of $\sim$ 280 $\mathrm{km \, s}^{-1}$ with widths of $\sim$ 30 $\mathrm{km \, s}^{-1}$.

The \ion{H}{i} profiles are broader (about 30 $\mathrm{km \, s}^{-1}$). Again, in region III the peak position is at lower velocities, but in regions I-1, I-2, and II we also find the peak position shifted to slightly lower velocities compared to the [\ion{C}{ii}] spectra.
\noindent
\begin{minipage}[t]{\textwidth}
\begin{threeparttable}
\centering 
\captionof{table}{Fluxes of [\ion{C}{ii}], \ion{H}{i}, and CO for the five SOFIA/GREAT pointing positions.\vspace{-10pt}}        
\begin{tabular}{c c c c c c c c c}          
\hline\hline                       
Region & RA (J2000) & DEC (J2000) & F([\ion{C}{ii}])$_{\mathrm{PACS}}$& F([\ion{C}{ii}])$_{\mathrm{GREAT}}$ & F(\ion{H}{i}) & F(CO(2-1)) & 
F(CO(1-0)) & $R_{\frac{2 \rightarrow 1}{1 \rightarrow 0}}$\\
&  & & [$10^{-16} \mathrm{W m}^{-2}$] & [$10^{-16}  \mathrm{W m}^{-2}$] & [Jy  $\mathrm{km \, s}^{-1}$] & [Jy  $\mathrm{km \, s}^{-1}$] & [Jy  $\mathrm{km \, s}^{-1}$] & \\
\hline 
I-1 & 12:15:38.48 & +36:19:32.4 & 3.98 $\pm$ 0.08 & 3.54 $\pm$ 0.49 & 0.48 $\pm$ 0.01 & 10.27 $\pm$ 2.21 & 1.77 $\pm$ 0.49 & 1.46 $\pm$ 0.52 \\
I-2 & 12:15:39.61 & +36:19:37.1 & 7.02 $\pm$ 0.12 & 9.62 $\pm$ 0.43 & 0.52 $\pm$ 0.01 & 16.64 $\pm$ 2.08 & 3.14 $\pm$ 0.34 & 1.33 $\pm$ 0.22  \\                            
I-3 & 12:15:40.74 & +36:19:34.3 & 4.85 $\pm$ 0.10 & 7.27 $\pm$ 0.40 & 0.55 $\pm$ 0.01 & 6.76 $\pm$ 1.95 & 1.12 $\pm$ 0.57 & 1.52 $\pm$ 0.89\\
II  & 12:15:40.83 & +36:19:04.7 & 5.03 $\pm$ 0.09 & 6.86 $\pm$ 0.32 & 0.49 $\pm$ 0.01 & 19.51 $\pm$ 2.34 & 6.76 $\pm$ 0.39 & 0.76 $\pm$ 0.10 \\
III & 12:15:36.52 & +36:20:01.7 & 1.18 $\pm$ 0.05 & 2.49 $\pm$ 0.62 & 0.58 $\pm$ 0.01 & 12.09 $\pm$ 2.73 & 3.69 $\pm$ 0.70 & 0.88 $\pm$ 0.25 \\
\hline                                             
\end{tabular}
\begin{tablenotes}
\item\small Notes.
The last column lists integrated line ratios of CO(2 $\rightarrow$ 1)/CO(1 $\rightarrow$ 0) with CO fluxes in K km s$^{-1}$. To convert from Jy $\mathrm{km \, s}^{-1}$ to $\mathrm{W m}^{-2}$, the fluxes should be multiplied by $a = 10^{-26} \frac{\nu_0}{c}$, where $a_{\ion{H}{i}} = 4.74 \times 10^{-23}$, $a_{\mathrm{CO(2\,\rightarrow\,1)}} = 7.69 \times 10^{-21}$ and $a_{\mathrm{CO(1\,\rightarrow\,0)}} = 3.85 \times 10^{-21}$.
\end{tablenotes} 
\label{tab:Fluxes} 
\end{threeparttable}

\includegraphics[width=0.97\textwidth]{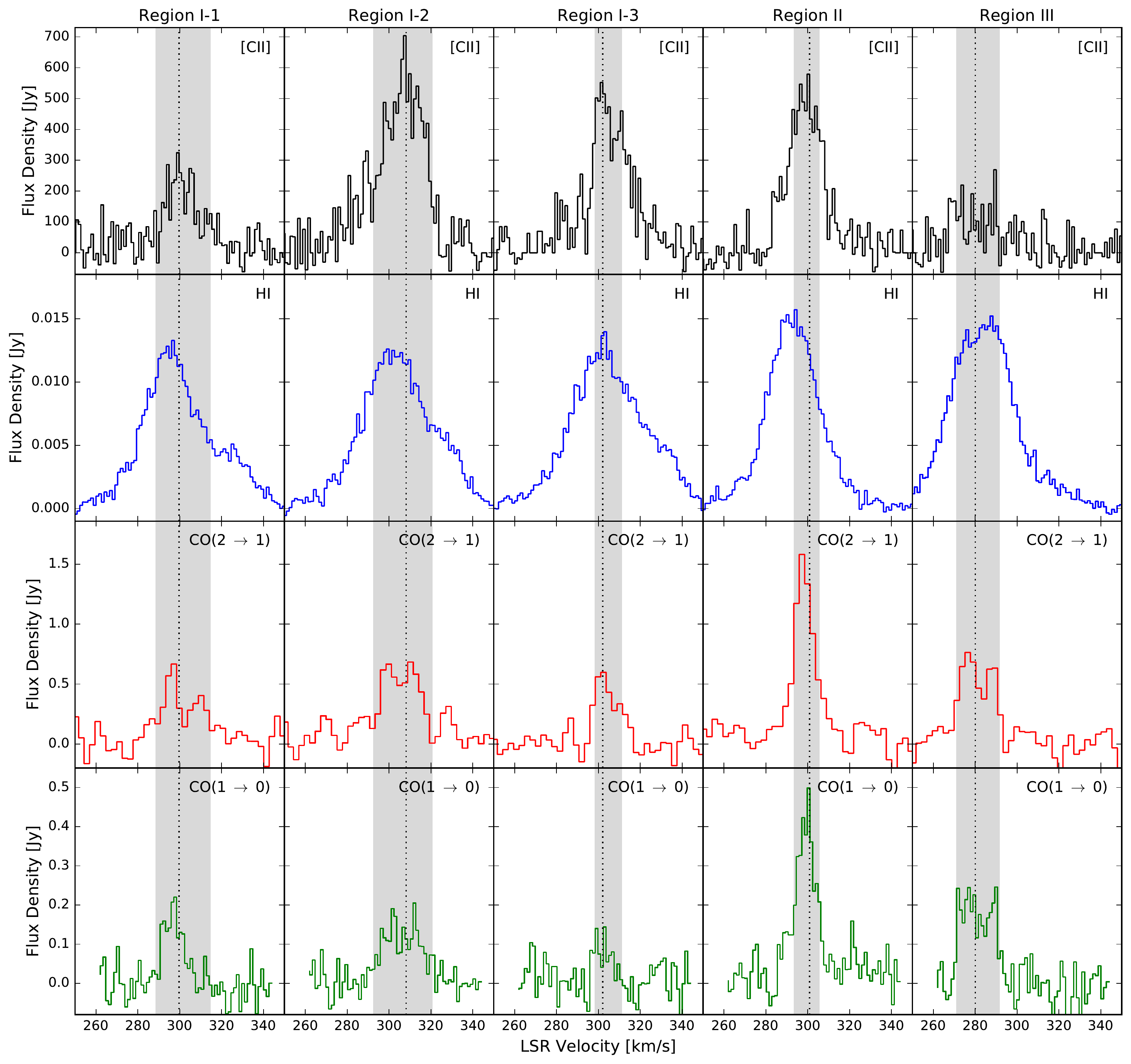} 
\captionof{figure}{Emission spectra of SOFIA/GREAT [\ion{C}{ii}] and ancillary CO and \ion{H}{i} in the five positions in NGC 4214. The spectra shown are based on the data that were spatially convolved with Gaussian kernels to the resolution of the SOFIA data (14\arcsec). The vertical lines indicate the channel with maximum [\ion{C}{ii}] emission, except for region III where it indicates the center of the broad profile. The gray-shaded areas indicate the FWHM ranges around the CO-peak positions. The original velocity resolutions shown here are 1.16 $\mathrm{km \, s}^{-1}$, 1.29 $\mathrm{km \, s}^{-1}$, 2.6 $\mathrm{km \, s}^{-1}$, and 1.30 $\mathrm{km \, s}^{-1}$ for [\ion{C}{ii}], \ion{H}{i}, CO(2 $\rightarrow$ 1), and CO(1 $\rightarrow$ 0), respectively.}
\label{fig:spectra}
\end{minipage}
\clearpage

We note that the CO emission in both transition lines is clearly strongest in region II with a width of $\sim$ 15 $\mathrm{km \, s}^{-1}$, while the other regions show widths between 20 and 25 $\mathrm{km \, s}^{-1}$. The spectra peak at $\sim$ 300 $\mathrm{km \, s}^{-1}$ except for region III where the velocity is again lower.
The spectra of CO(2 $\rightarrow$ 1) and CO(1 $\rightarrow$ 0) have similar shapes. However, in regions I-1, I-2, and I-3 there are some features to the right and left of the central peak that are only visible in one of the two datasets.

Table \ref{tab:Fluxes} shows that the \ion{H}{i} fluxes are almost constant throughout the five regions, as is also indicated by the spectra. We find that the CO emission is weak compared to [\ion{C}{ii}] with flux ratios of [\ion{C}{ii}]/CO(1 $\rightarrow$ 0) between 2 $\times$ 10$^4$ and 17 $\times$ 10$^4$ with a clear minimum in region III and the maximum in region I-2 (see also \citealt{Cormier2010}).

\section{\texorpdfstring{Decomposition of [\ion{C}{ii}] spectra}{Decomposition of [CII] spectra}}
\label{sec:Decomp}
\begin{figure*}
\includegraphics[width=1.\textwidth]{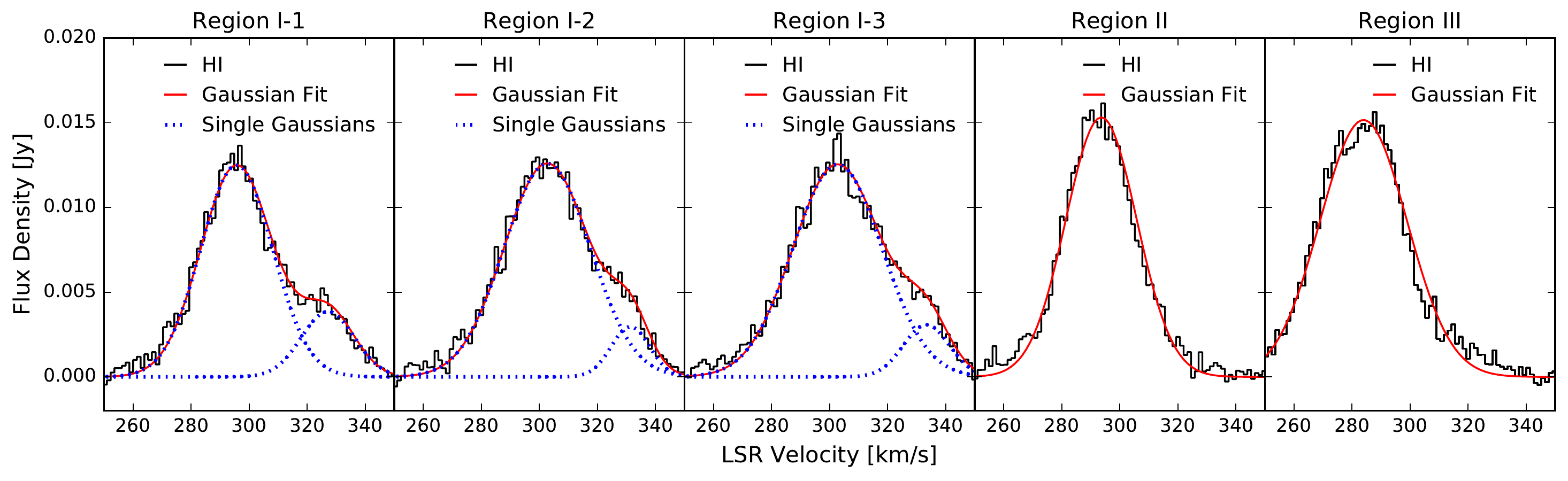}
\includegraphics[width=1.\textwidth]{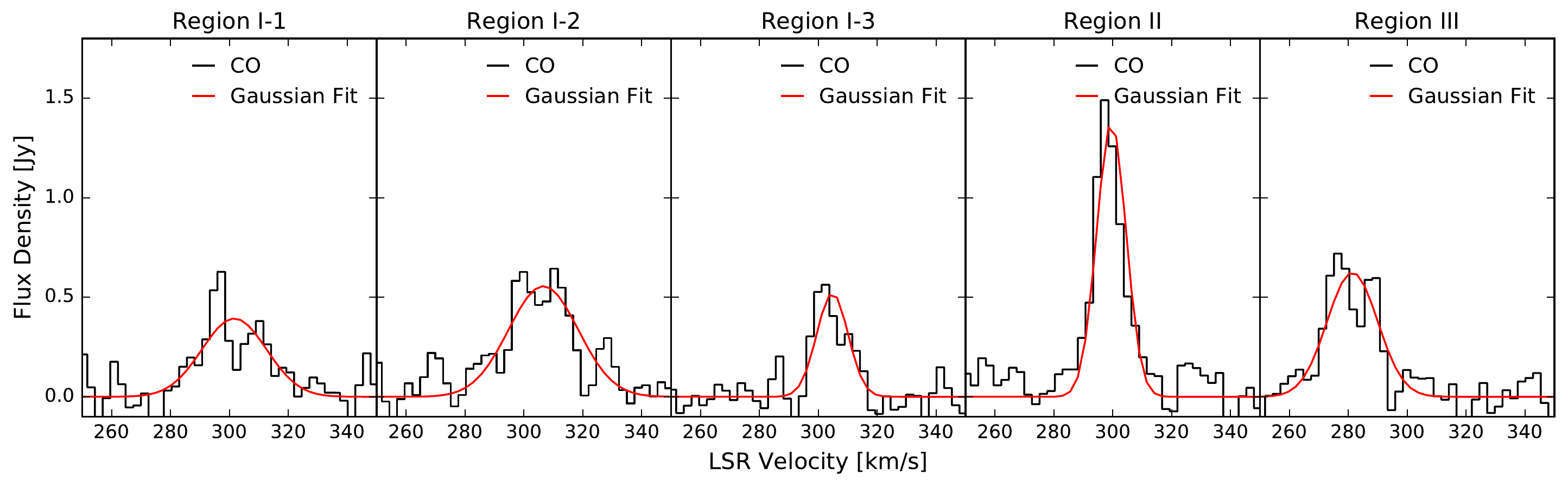}
\caption{Upper row: \ion{H}{i} spectra of the five pointings in NGC 4214 fitted with Gaussian curves. For the spectra in region I, we use two Gaussians (single components in blue). Bottom row: CO(2\,$\rightarrow$\,1) fitted with single Gaussian curves.}
\label{fig:HIfit}
\end{figure*}

\begin{table*}
\centering
\begin{threeparttable}
\caption{Parameters of the Gaussian fitting of the \ion{H}{i} and CO spectra.}
\centering
\begin{tabular}{l c c c c c c} \hline \hline
& \ion{H}{i} & & & & CO(2\,$\rightarrow$\,1) & \\ 
Region & $p_{\mathrm{HI},1}$ & FWHM$_{\mathrm{HI},1}$ & $p_{\mathrm{HI},2}$ & FWHM$_{\mathrm{HI},2}$ & $p_{\mathrm{CO}}$ & FWHM$_{\mathrm{CO}}$ \\
 & [$\mathrm{km \, s}^{-1}$] & [$\mathrm{km \, s}^{-1}$] & [$\mathrm{km \, s}^{-1}$] & [$\mathrm{km \, s}^{-1}$] & [$\mathrm{km \, s}^{-1}$] & [$\mathrm{km \, s}^{-1}$]\\
\hline
I-1 & 295.7 & 29.1 & 327.1 & 21.5 & 301.6 & 25.8\\
I-2 & 302.7 & 35.0 & 331.4 & 15.5 & 306.6 & 27.8 \\
I-3 & 302.7 & 36.2 & 333.4 & 18.7 & 304.8 & 12.5 \\
II & 293.6 & 28.7 & - & - & 299.5 & 11.7 \\
III & 284.0 & 35.3 & - & - & 281.4 & 20.2 \\ \hline
\end{tabular}
\begin{tablenotes}
\item\small Notes. Peak positions $p$ and FWHMs (= 2.35 $\sigma$). Two Gaussians were fitted to the \ion{H}{i} spectra of the pointings in region I.
\end{tablenotes}
\label{tab:gaussianfits}
\end{threeparttable}
\end{table*}
The main goal of this work is to determine how much of the [\ion{C}{ii}] emission traces the dense molecular material that is also probed by CO. If it were a one-to-one tracer, the velocity profiles of [\ion{C}{ii}] and CO would be similar. However, as Figure \ref{fig:spectra} already shows, the [\ion{C}{ii}] profiles in NGC 4214 show broader wings that are not seen in CO. To determine the maximum amount of [\ion{C}{ii}] that can be associated with the molecular phase and to investigate the origin of these wings, we decompose the spectrally resolved [\ion{C}{ii}] observations into contributions correlated with H$_2$ (as traced by CO(2\,$\rightarrow$\,1)) and \ion{H}{i}. In doing so, we assume that CO(2\,$\rightarrow$\,1) and \ion{H}{i} trace material with uniform conditions. The consideration of atomic gas with different conditions is discussed in Section \ref{sec:atomicgas}.

We decompose our [\ion{C}{ii}] spectra into a linear combination of \ion{H}{i} and CO in two different ways. The first approach (channel method) uses the CO and \ion{H}{i} spectra directly by adding them manually (see Appendix \ref{sec:appendix}). The second approach uses Gaussian fits for the decomposition, as described in the following. Although neither method shows striking inconsistencies, we focus on the Gaussian fit method as it appears to be more robust.

\subsection{Description of the Gaussian fit method}
We use Gaussian fits to decompose the [\ion{C}{ii}] spectra. In this approach, we first fit \ion{H}{i} and CO(2\,$\rightarrow$\,1) spectra independently with Gaussians. We fit two independent Gaussian curves to the \ion{H}{i} spectra in region I as there seem to be two distinct components (see Sect. \ref{fig:spectra}). For CO(2\,$\rightarrow$\,1), we only use one component because of the lower signal-to-noise ratio. Table \ref{tab:gaussianfits} lists the fitted peak positions and widths (FWHM of the Gaussian curves) for the \ion{H}{i} and CO spectra in the five regions in NGC 4214. Figure \ref{fig:HIfit} shows that two Gaussians reproduce the \ion{H}{i} data well in the pointings of region I; however, the parameters of the second Gaussian are less well constrained. We note that the first component of \ion{H}{i} in region I has the same width as the \ion{H}{i} spectra in region II and III indicating that the second component might be emission on the line of sight not connected to the star-forming region. 
The CO(2\,$\rightarrow$\,1) spectrum in region I-1 is fitted with a very broad Gaussian (FWHM $\approx$ 26 $\mathrm{km \, s}^{-1}$), whereas the CO(1\,$\rightarrow$\,0) is fitted with a FWHM of $\approx$ 11 $\mathrm{km \, s}^{-1}$. We approximate the CO width in region I-1 for the latter analysis to be 18 $\mathrm{km \, s}^{-1}$, the average FWHM of the original fitting of the CO(2\,$\rightarrow$\,1) and CO(1\,$\rightarrow$\,0) spectra. We discuss how changing the FWHM in this region to the individually fitted values (11 and 26 $\mathrm{km \, s}^{-1}$) affects the decomposition results in the following section. In the other regions fitting the CO(1\,$\rightarrow$\,0) and CO(2\,$\rightarrow$\,1) spectra yields comparable widths.

We then normalize the [\ion{C}{ii}] spectra to their maxima and fit them with Gaussians of the form
\begin{equation}
\begin{split}
g(\mathrm{v}) = & \, A_{1} \, \mathrm{exp}\left(-\frac{(\mathrm{v} - p_{\mathrm{HI},1})^2}{2 \sigma^{2}_{\mathrm{HI},1}}\right) + A_{2} \, \mathrm{exp}\left(-\frac{(\mathrm{v} - p_{\mathrm{HI},2})^2}{2 \sigma^{2}_{\mathrm{HI},2}}\right) \\
& + B \, \mathrm{exp}\left(-\frac{(\mathrm{v} - p_{\mathrm{CO}})^2}{2 \sigma^{2}_{\mathrm{CO}}}\right).
\end{split}
\label{eq:twogauss}
\end{equation}

The peak positions $p$ and widths $\sigma$ are taken as fixed parameters determined by the above-mentioned fits. The coefficients $A_i$ (\ion{H}{i} amplitudes) and $B$ (CO) are used as free parameters for the fit. In region II and III, we use only one \ion{H}{i} term. In region III, we rebin the [\ion{C}{ii}] spectrum to a resolution of 5 $\mathrm{km \, s}^{-1}$ to improve the signal-to-noise ratio. 

\subsection{Results}
\label{sect:decompresults}
\begin{table*}
\centering
\begin{threeparttable}
\caption{Results from the Gaussian fit decomposition.}
\begin{tabular}{l c c c c c} \hline \hline
Region & $A_{1}$ & $A_{2}$ & $B$ & $I_{[\ion{C}{ii}]}(\mathrm{CO})$/$I_{[\ion{C}{ii}]}$ & \\ 
& \ion{H}{i} & \ion{H}{i} & CO & central value & 1$\sigma$ interval\\ \hline
I-1 & 0.12 $\pm$ 0.08 & 0.06 $\pm$ 0.05 & 0.62 $\pm$ 0.10 & 77.4\% & 65\% - 85\%\\
I-2 & 0.04 $\pm$ 0.10 & 0.00 $\pm$ 0.04 & 0.70 $\pm$ 0.11 & 93.0\% & 85\%-100\%\\
I-3 & 0.35 $\pm$ 0.04 & 0.04 $\pm$ 0.04 & 0.56 $\pm$ 0.08 & 34.3\% & 25\%-45\% \\
II & 0.29 $\pm$ 0.04 & - & 0.71 $\pm$ 0.06 & 49.5\% & 40\%-60\%\\
III & 0.29 $\pm$ 0.11 & - & 0.18 $\pm$ 0.24 & 17.5\% & 0\%-30\%\\
\hline
\end{tabular}
\begin{tablenotes}
\item\small Notes. $A_{1}$, $A_{2}$ and $B$ are the coefficients of the components (Eq. \eqref{eq:twogauss}). Columns 5 and 6 list the central values and 1$\sigma$ intervals of the fraction of [\ion{C}{ii}] intensity associated with CO (see Sect. \ref{sect:decompresults}). The intensity fractions associated with \ion{H}{i} are the remaining fraction of the [\ion{C}{ii}] intensity.
\end{tablenotes}
\label{tab:fitresults}
\end{threeparttable}
\end{table*}

\begin{figure*}
\includegraphics[width=1.\textwidth]{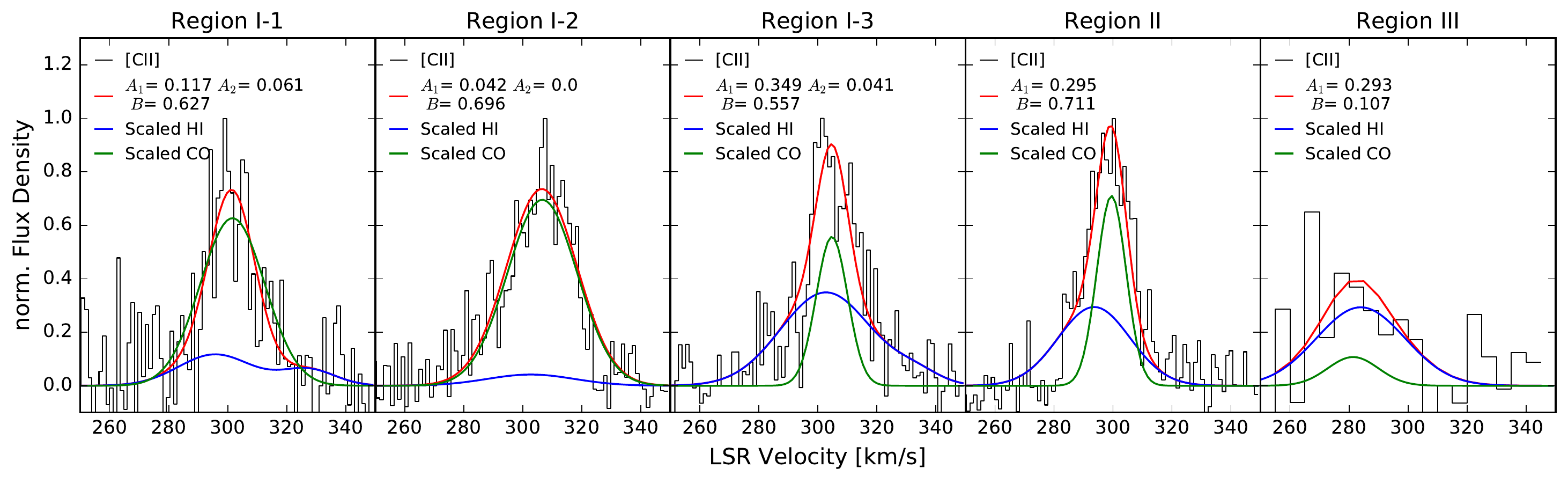} 
\caption{Gaussian fit method. [\ion{C}{ii}] spectra in the five pointings fitted with Gaussian curves as described by Eq. \eqref{eq:twogauss}. In region II and III we only fit two Gaussian curves, but in the pointings of region I we use two \ion{H}{i} components. The [\ion{C}{ii}] spectra in region III was rebinned to 5 $\mathrm{km \, s}^{-1}$.}
\label{fig:CIIgaussfit}
\end{figure*}

Figure \ref{fig:CIIgaussfit} shows the [\ion{C}{ii}] spectra in black with the best fitting curves overlaid in red. The fitted amplitudes and their uncertainties are given in Table \ref{tab:fitresults}. The stated uncertainties are the 1$\sigma$ errors from the fitting procedure. In the three pointings of region I, we find that the best fit is provided when the contribution of the second \ion{H}{i} component is close to zero. This supports our assumption that this second component may not be connected to the star-forming region while the [\ion{C}{ii}] emission is.

In addition, we note that the CO amplitudes are higher than the sum of \ion{H}{i} amplitudes in all regions except for region III, where the uncertainties are high. The [\ion{C}{ii}] spectra are best decomposed when using large fractions of the CO(2\,$\rightarrow$\,1) spectra in agreement with the scaling coefficients used in the channel method (Appendix \ref{sec:appendix}). 
To verify that the Gaussian fits recover the [\ion{C}{ii}] fluxes, we compare the velocity integrated fluxes of the original [\ion{C}{ii}] data with the combined Gaussians using the same velocity ranges as before. The ratios of Gaussians and direct integration are very close to unity and therefore our fits recover all of the [\ion{C}{ii}] emission. 

With the spectral decomposition of [\ion{C}{ii}] emission determined in this way, we calculate the fractions of [\ion{C}{ii}] intensity associated with CO(2\,$\rightarrow$\,1) and \ion{H}{i}. Each fraction is determined as the ratio between the integrated [\ion{C}{ii}]-fitted profiles and the accordingly scaled \ion{H}{i} and CO spectra that were used in the decomposition. Simply put, we compare the areas below the red curves (final fit) in Figure \ref{fig:CIIgaussfit} to the areas below the green and blue curves (corresponding \ion{H}{i} and CO contributions to the final fit). The [\ion{C}{ii}] intensity fractions associated with CO(2\,$\rightarrow$\,1) are listed in Table \ref{tab:fitresults}. In region I-2, we find almost no contribution from the \ion{H}{i}, and the CO(2\,$\rightarrow$\,1) contribution to the [\ion{C}{ii}] flux is also higher than the \ion{H}{i} contribution in region I-1. In region II, the parts are almost equally split. In region III, we can only associate a rather small fraction of CO emission with the [\ion{C}{ii}].

We validate the uncertainties on the amplitudes given by the IDL routine GAUSSFIT in the following manner. We vary the CO amplitude from 0.3 to 0.9 in steps of 0.1 and re-fit the \ion{H}{i} amplitude. For each chosen CO amplitude, we estimate the goodness of the fit by considering the residuals (signal $-$ fit) normalized by the noise level of the [\ion{C}{ii}] data. We calculate the $\chi^2$ by summing these residuals over the channels between 280 and 320 km/s. The 1$\sigma$ interval that we find corresponds well with the uncertainties given by the fitting procedure. This range of amplitudes is then translated into uncertainties on the intensity fractions (column 5 of Table \ref{tab:fitresults}). We find that these uncertainties are about $\pm$ 10\%, except for region III where they are at least $\pm$ 20\%.
For region I-1, we also explore the sensitivity of the results to the value
of the CO FWHM because the two available CO lines yield substantially different values. When using 26\,km s$^{-1}$, i.e., the value derived from the CO(2\,$\rightarrow$\,1), we obtain a poor fitting result because the CO(2\,$\rightarrow$\,1) line is broader than the observed [\ion{C}{ii}] line.  As a result, the wings and the peak in the [\ion{C}{ii}] cannot be reproduced at the same time. When using the FWHM determined from CO(1\,$\rightarrow$\,0), we obtain a reasonable decomposition with a CO amplitude of 0.55 and a residual comparable to when using the average FWHM. This results in a CO-associated intensity fraction of (56$\pm$10)\%, a value around 20\% lower than that given in Table \ref{tab:fitresults} calculated with the average FWHM (18 km s$^{-1}$). We note that the maximum CO percentage provides a hard limit on the amount of [\ion{C}{ii}] that can be associated with CO. The lower limit is less strict because it depends on the nature of the \ion{H}{i}-emitting gas (see Sect. \ref{sec:atomicgas}). The \ion{H}{i} profiles might contain narrow and broad components that we cannot separate owing to the limited spatial resolution of the \ion{H}{i} observations.

The quality of the input spectra that define the Gaussian components plays an important role in our method. The uncertainties of our results could be further reduced with deeper CO observations. This is particularly true for region I-1, but is true for the other regions as well.

\section{Gas masses from different tracers}
\label{sec:AssociatedGas}

\subsection{\texorpdfstring{C$^+$ column densities and associated gas masses}{C+ column densities and associated gas masses}}
The decomposition described in the previous section divides the [\ion{C}{ii}] emission into fractions associated with \ion{H}{i} and CO(2\,$\rightarrow$\,1) emission. We find that the CO fraction tends to be larger; however, the [\ion{C}{ii}] profiles show both narrow and broad components.
Assuming different sets of conditions in the ISM, we calculate the gas masses that these components would correspond to. In principle, at our spatial resolution (14\arcsec $\sim$ 200pc), the ISM should contain ionized gas, cold neutral medium (CNM), warm neutral medium (WNM), and dense neutral/molecular gas (PDR). While the narrow components in the [\ion{C}{ii}] profiles can be associated with the PDR, the broad components can correspond to ionized gas, WNM, and/or CNM. 

We neglect the contribution from ionized gas because of a previous study of NGC 4214 by \cite{Dimaratos2015}. They perfomed a self-consistent radiative transfer modeling of the \ion{H}{ii} gas and PDR in regions I and II of NGC 4214 using a suite of mid-IR and far-IR lines. They find that the \ion{H}{ii} region models contribute only a small percent of the [\ion{C}{ii}] emission. Using the [\ion{N}{ii}] lines at 122 $\mu$m and 205 $\mu$m, they also estimate the contribution of a diffuse ionized phase to the [\ion{C}{ii}] emission to be less then 8\%.

\begin{table*}
\centering
\begin{threeparttable}
\caption{Assumed sets of conditions for ISM phases used to calculate [\ion{C}{ii}]-associated gas masses, and observed \ion{H}{i} and H$_2$ masses.}
\vspace{-10pt}
\begin{tabular}{c c c | c c c c} 
 & & & CNM & WNM & PDR CO \\ \cline{3-6}
 & & Collision partner of C$^+$ & H & H & H$_2$ \\
 & & T [K] & 80 & 8000 & 150 \\
 & & $n$ [cm$^{-3}$] & 100 & 0.5 & 10$^4$ \\
 & & $n_{\mathrm{crit}}$ [cm$^{-3}$] & 3000 & 1600 & 6100 \\
\cline{3-6}
& & & & & \\
\hline \hline

Region & $\mathrm{M}_{\mathrm{HI}}$ & $ \mathrm{M}_{\mathrm{H}_2}$ & $\mathrm{M}_{\mathrm{gas}}$([\ion{C}{ii}]$_{\mathrm{\ion{H}{i}, \, CNM}}$) & $\mathrm{M}_{\mathrm{gas}}$([\ion{C}{ii}]$_{\mathrm{\ion{H}{i}, \, WNM}}$)&  $\mathrm{M}_{\mathrm{gas}}$([\ion{C}{ii}]$_{\mathrm{CO, \, PDR}}$)\\ 
 & [$10^5 \, \mathrm{M}_{\sun}$] & [$10^5 \, \mathrm{M}_{\sun}$]& [$10^5 \, \mathrm{M}_{\sun}$]  & [$10^7 \, \mathrm{M}_{\sun}$]  & [$10^5 \, \mathrm{M}_{\sun}$] \\ \hline \hline
 I-1 & 9.30 & 1.94 & 23.16$\substack{+9.01 \\ -9.37}$ & 7.57$\substack{+2.95 \\ -3.07}$ & 6.88$\substack{+0.94 \\ -0.90}$\\
 I-2 & 10.11 & 3.12 & 4.25$\substack{+20.73 \\ -4.25}$ & 1.39$\substack{+6.78 \\ -1.39}$ & 24.58$\substack{+0.43 \\ -2.08}$ \\
 I-3 & 10.62  & 1.27 & 117.39$\substack{+14.72 \\ -23.03}$ & 38.40$\substack{+4.82 \\ -7.53}$ & 7.14$\substack{+2.30 \\ -1.47}$ \\ 
 II & 9.36 & 3.65 & 60.36$\substack{+14.47 \\ -10.46}$ & 19.74$\substack{+4.73 \\ -3.43}$ & 6.44$\substack{+1.05 \\ -1.45}$ \\ 
 III & 11.20 & 2.42 & 53.33$\substack{+11.31 \\ -8.08}$ & 17.44$\substack{+3.70 \\ -2.64}$ & 1.13$\substack{+0.81 \\ -1.13}$ \\ 
  \hline \hline\\

\end{tabular}
\begin{tablenotes}
\item\small Notes. We present the observed \ion{H}{i} and H$_2$ masses and compare them to theoretical values of the gas mass calculated from the \ion{H}{i}- or H$_2$-associated components of the [\ion{C}{ii}] intensities under the assumption that all of the \ion{H}{i}- or CO-associated [\ion{C}{ii}] emits from an ISM phase with the conditions stated above. The uncertainties result from the uncertainties given in Table \ref{tab:fitresults}. PDR conditions are from \cite{Dimaratos2015}, critical densities from \cite{Goldsmith2012}.
\end{tablenotes}
\label{tab:GasMasses}
\end{threeparttable}
\end{table*}

We study the extended component, i.e., the \ion{H}{i}-associated [\ion{C}{ii}] flux, as well as the CO-matched dense material by calculating gas masses (corresponding to \ion{H}{i} or H$_2$ masses) assuming different sets of ISM conditions (see Table \ref{tab:GasMasses}). None of the quoted masses includes helium. 

We calculate C$^+$ column densities from the observed intensity $I_{\mathrm{[CII]}}$ following \cite{Madden1997}:
\begin{equation}
\begin{split}
 N(\mathrm{C}^+) & = 1.06 \cdot 10^{15} \, \frac{\int F_{\nu}\D v \, [\mathrm{Jy \, km\,s^{-1}}]}{\Theta['']^2} \\
 & \times \, \left(\frac{1 + 2 \, \mathrm{exp}(-91.3 \,\mathrm{K}/T) + n_{\mathrm{crit}}/n}{2 \, \mathrm{exp}(-91.3 \, \mathrm{K}/T)}\right).
 \label{eq:CIIcolumn}\\
 \end{split}
\end{equation}
Here $T$ is the kinetic gas temperature in Kelvin, and $n$ the density and $n_{\mathrm{crit}}$ the critical density of the main collision partner of C$^+$ both in cm$^{-3}$.
We assume that all of the carbon is ionized in the CNM, WNM, and in the PDR, and that the [\ion{C}{ii}] line is excited by collisions with atomic hydrogen in the phases associated with \ion{H}{i} emission and by collisions with molecular hydrogen in the PDR (see Table \ref{tab:GasMasses}). We use the intensity fractions from Table \ref{tab:fitresults} to scale the column density according to the ISM phase that is considered.

For the abundance of ionized carbon, $X_{\mathrm{C}^+/\mathrm{H}} = 5.07 \cdot 10^{-5}$ is used in region I and  $X_{\mathrm{C}^+/\mathrm{H}} = 7.24 \cdot 10^{-5}$ is used in region II \citep{Kobulnicky1996}. There is no measurement for region III, so we used the lower value since this region is thought to be less evolved.

We then calculate the gas masses traced by [\ion{C}{ii}] and associated with CO and \ion{H}{i} by adapting equation (3) from \cite{Perez-Beaupuits2015} to our data. We assume a beam filling factor of 1:
\begin{equation}
 \mathrm{M}_{\mathrm{gas}}[\mathrm{M}_{\odot}] = 3.16 \times 10^{-16} \, \frac{N(\mathrm{C}^+)}{X_{\mathrm{C}^+/\mathrm{H}}}.
 \label{eq:CIImass}
\end{equation}

The resulting gas masses are reported in Table \ref{tab:GasMasses} using intensity fractions with uncertainties from Table \ref{tab:fitresults}. We emphasize that these masses are theoretical values assuming that the \ion{H}{i} and CO-associated fractions of the [\ion{C}{ii}] intensity originate from regions of uniform ISM conditions. We discuss these model values further in Section \ref{sec:Discussion}. The gas masses are sensitive to the assumed ISM conditions. For example, changing the CNM temperature from 80 to 100 K results in CNM-associated gas masses that are reduced by 20\% compared to the values given in Table \ref{tab:GasMasses}. By changing the electron density from 100 to 200 cm$^{-3}$, the gas mass is about half of the reported value. Similarly, changing the PDR temperature to 100\,K, the theoretical mass is reduced by 18\%. By decreasing the assumed PDR density by a factor of 2 to 5000 cm$^{-3}$, the resulting mass is reduced by 20\%. However, owing to the inherently high density in this phase, the masses are less sensitive to an increase in the density. If the density is increased to 10$^5$\,cm$^{-3}$, the mass increases by 25\%. We discuss the sensitivity to the assumed parameters further in Section \ref{sec:gas_discussion}.

\subsection{\texorpdfstring{Gas masses traced by \ion{H}{i} and CO}{Gas masses traced by HI and CO}}
\label{sec:Methods_GasMasses}

In addition to the gas masses traced by [\ion{C}{ii}], we also calculate the atomic and molecular gas masses using the \ion{H}{i} and CO(1\,$\rightarrow$\,0) intensities directly (Table \ref{tab:GasMasses}).

The \ion{H}{i} mass is calculated using the relation
\begin{equation}
\begin{split}
 \mathrm{M}_{\mathrm{HI}}[\mathrm{M}_{\odot}] = 2.36 \times 10^{5} \, (D[\mathrm{Mpc}])^2 \, \int F_{\nu} \D v \, [\mathrm{Jy \, km\,s^{-1}}].
\end{split}
 \label{eq:HImass}
\end{equation}

To calculate the molecular gas mass from CO(1\,$\rightarrow$\,0) emission, we use equation (2) from \cite{Bolatto2013}
\begin{equation}
\begin{split}
 \mathrm{M}_{\mathrm{H}_2}[\mathrm{M}_{\odot}] &= \frac{\alpha_{\mathrm{CO}}}{1.36}\,L_{\mathrm{CO(1\rightarrow0)}},\\
 \end{split}
\label{eq:H2mass}
\end{equation}
where $\alpha_{\mathrm{CO}} = 4.3\,\mathrm{M}_{\odot} (\mathrm{K\,km\,s^{-1}\,pc^{2}})^{-1}$ is the Galactic value of the CO-to-H$_2$ conversion factor and $L_{\mathrm{CO(1\rightarrow0)}}$ is expressed in K km s$^{-1}$ pc$^2$. The factor of 1.36 is the helium contribution that we do not consider here.
Table \ref{tab:GasMasses} lists the \ion{H}{i} and H$_2$ masses calculated from \ion{H}{i} and CO emission directly.

While the \ion{H}{i} mass distribution is almost constant throughout the regions, the values for H$_2$ masses differ from region to region by up to a factor of 2; the maximum is found in region II.
We find ratios of $ \mathrm{M}_{\mathrm{H}_2}$ to $\mathrm{M}_{\mathrm{HI}}$ of $\sim$ 0.2. These ratios are higher than the 0.02 found by \cite{HeraclesPaper} for the whole galaxy because we only study the star-forming regions and therefore only include small fractions of the extended \ion{H}{i}.

\section{Discussion}
\label{sec:Discussion}
\subsection{\texorpdfstring{[\ion{C}{ii}] emission from different ISM phases}{[CII] emission from different ISM phases}}
\label{sec:gas_discussion}
We showed in our analysis that the [\ion{C}{ii}] velocity profiles do not have the same shapes as the CO profiles, but they show broad wings that can be reproduced by \ion{H}{i}. We want to understand what this means in terms of how reliably [\ion{C}{ii}] traces the molecular phase and what the origin of these broad wings might be. With our decomposition and the calculated model gas masses (Table \ref{tab:GasMasses}), we can explore this issue.

\subsubsection{Molecular gas}
\label{sec:moleculargas}
The [\ion{C}{ii}] gas mass associated with the dense PDR (traced by CO emission) reaches its maximum in region I-2 and its minimum in region III. In the other regions the mass is almost constant. 
Except for region III, we find that the [\ion{C}{ii}] emission traces 3-8 times more mass than CO. 

We can estimate how much of the molecular mass is CO-dark by comparing the molecular gas mass found with [\ion{C}{ii}] emission assuming PDR conditions with the molecular gas mass directly determined from CO(1\,$\rightarrow$\,0) observations (column 6 and 3 in Table \ref{tab:GasMasses}, respectively). We calculate the CO-dark molecular mass fraction as the molecular mass derived from [\ion{C}{ii}] divided by the total molecular mass (from CO and [\ion{C}{ii}]). In regions I-1, I-2, and I-3, the fraction of total molecular gas mass only seen by [\ion{C}{ii}] is high with (78$\pm$13)\%, (89$\pm$10)\%, and (85$\substack{+15 \\ -25}$)\%, respectively, suggesting a large amount of CO-dark-H$_2$ gas extended over region I as is expected in low-metallicity environments. The errors quoted here result from the 1$\sigma$ uncertainties in the decomposition.
Since \cite{MaApell1999} and \cite{MacKenty2000} found that region I is resolved into two different regions around two massive star clusters, it is not surprising that we find different fractions in the three pointings. We note, that region I-2 covers the super star cluster (cluster A, see \citealt{MaApell1999} or \citealt{MaizApellaniz2001}), while a second cluster (labeled B) is covered by pointing I-3. While region I has an age of 3 - 4 Myr \citep{Ubeda2007Results}, region II is a more compact and younger ($\sim$ 2 Myr) star-forming region with higher metallicity (by $\sim$ 0.1 dex; \citealt{Kobulnicky1996}. In region II, we report that (64$\pm$20)\% of the total molecular mass is found with [\ion{C}{ii}] emission. This value is lower than the those found in region I and might be explained by increased dust shielding, as indicated by the slightly higher metallicity and less porous ISM structure of this younger region. On average, around 79\% of the total molecular mass is CO-dark.

The gas mass found in CO slightly exceeds the mass found by [\ion{C}{ii}] in region III. We can therefore make no statement about the CO-dark gas in this region owing to a combination of an uncertain decomposition and low [\ion{C}{ii}] intensity. Region III is the least evolved region with no visible star formation.

We note that the [\ion{C}{ii}]-associated gas masses vary with the assumed conditions. While small deviations from the assumed ISM conditions could reduce the CO-dark fractions, we find average PDR densities of $\sim 500$ cm$^{-3}$ or temperatures of $\sim 35$ K are needed to explain the observed H$_2$ masses without any CO-dark gas. Those conditions are not in agreement with the PDR analysis of \cite{Dimaratos2015}. These are strong arguments in favor of a massive CO-dark gas phase.

Instead of masses, we can also quote fractions of total [\ion{C}{ii}] intensity assigned to the CO-dark phase. In region II, we can assign (33$\pm$9)\% of the total [\ion{C}{ii}] intensity to this phase. In regions I-1, I-2, and I-3 we find values of (58$\pm$11)\%, (87$\pm$12)\%, and (32$\pm$12)\%, respectively. 

\subsubsection{Atomic gas}
\label{sec:atomicgas}
For the atomic gas, we have estimated gas masses assuming WNM and CNM conditions.
In the WNM case, very large masses, on the order of 10$^8$ M$_\odot$, would be required in order to explain the observed [\ion{C}{ii}] emission. Given that these are much larger than the observed \ion{H}{i} masses from the 21 cm line, only $\sim$1\% of the observed [\ion{C}{ii}] intensity can arise from the WNM.
In their study of the Milky Way, \cite{GOTCCII} also find no significant contribution of the WNM to the [\ion{C}{ii}] emission.

Nonetheless, the WNM can make up a significant fraction of the total atomic content (e.g., \citealt{Kulkarni1987, Heiles2003b}), especially at large velocities due to high temperatures. The widths of the \ion{H}{i} profiles that we observe could therefore be caused by the WNM. In principle, it might be better to first subtract any WNM contribution to the \ion{H}{i} line as the [\ion{C}{ii}] line is less sensitive to this warm phase. However, it is not easy to determine what causes the broad \ion{H}{i} profile.
\cite{Sellwood1999} and \cite{Tamburro2009} argue that  turbulence stirred by magnetorotational instability can also lead to broad line widths. Moreover, there could also be significant substructure within the telescope beam. To constrain any contribution from the WNM, further \ion{H}{i} observations with higher spatial resolution would be needed.

Even if we assume that the \ion{H}{i}-associated [\ion{C}{ii}] originates from the CNM, the corresponding masses exceed the upper limits given by the \ion{H}{i} observations directly. We find that between 4 and 10\% of our total [\ion{C}{ii}] intensity can arise from the CNM in order to reproduce the \ion{H}{i} 21 cm masses.
In region III, we can assign (17$\pm$6)\% of the [\ion{C}{ii}] emission to the CNM. In a study of the metal-poor irregular galaxy IC10, \cite{Madden1997} assigned about 10\% of [\ion{C}{ii}] emission to the CNM (and $\sim$ 10\% to the ionized medium) in agreement with our results. This indicates that the \ion{H}{i}-associated [\ion{C}{ii}] not only emits from the CNM, but also from a denser phase. To match the \ion{H}{i} masses, we would need densities of about 1000 cm$^{-3}$ on average when assuming temperatures of 80 K and a critical density of 3000 cm$^{-3}$. We note that it is not realistic that all the \ion{H}{i}-associated [\ion{C}{ii}] emits under these conditions, but it indicates that our CNM is not dense enough to be the sole origin of the broad wing component in the [\ion{C}{ii}] profiles. Furthermore, such high densities are unlikely as they would presumably appear as narrow wings, and not the broad wings that are seen in the [\ion{C}{ii}]. The true nature of these wings is not clearly understood.

Parts of the \ion{H}{i}-associated [\ion{C}{ii}] might come from the same dense PDR regions that are also traced by CO. This dense component of the \ion{H}{i} profile should have a similar velocity profile to CO, and as Figure \ref{fig:spectra} shows, it would be possible to fit a dense CO-associated component to the \ion{H}{i} near the peak. Therefore, when using only the three tracers discussed in this work, it becomes difficult to separate dense and extended components. Our findings show that further studies with additional, clearly separable tracers of molecular and atomic gas phases would be needed to resolve this issue. 

\subsection{Comparison with other sources}
Previous studies on star-forming regions and ISM conditions in nearby galaxies have shown a complex origin of [\ion{C}{ii}] emission. \cite{Requena-Torres2016} performed a study of carbon gas with high spatial and velocity resolution in low-metallicity star-forming regions N 66, N 25 + 26, and N 88 of the Small Magellanic Cloud (SMC). While N 66 is an extended (100 - 150 pc) gas cloud with bright core, N 25 and N 26 are part of a group of \ion{H}{ii} regions and N 88 is compact ($\sim$ 1 pc) source in the SMC wing. They found [\ion{C}{ii}] emission profiles that are up to 50\% wider in velocity than the corresponding CO profiles. They suggest that the gas traced with CO is embedded in a larger molecular cloud only seen by [\ion{C}{ii}] in the observed region. From studies on the C$^+$ column densities they concluded that most of the [\ion{C}{ii}] emission originates from CO-dark molecular gas. Only a few percent are contributed by either ionized and atomic hydrogen. The transition lines of CO only trace between 5 an 40\% percent of the molecular gas found in the observed regions.

Like the SMC, the LMC provides spatially and velocity-resolved observations of low-metallicity environments. \cite{Okada2015} studied the star-forming region N 159 in the LMC with [\ion{C}{ii}] SOFIA/GREAT observations  in comparison with different CO transition lines at a spatial resolution of $\sim$ 4 pc.  With a similar approach using Gaussian curves to fit the [\ion{C}{ii}] lines they found that the fraction of the [\ion{C}{ii}] that cannot be attributed to the gas traced by CO is 20\,\% around the regions with maxium CO emission and up to 50\% in the area between the CO cores. Additionally, they estimate the contribution from ionized gas to be $\le 15$\% in the whole observed region. While these results were obtained with a [\ion{C}{ii}] map, we only studied regions with strong CO emission and find that the fraction of [\ion{C}{ii}] emission associated with CO differs greatly from region to region. Averaged over all regions we can associate $\sim$ 54\% with the CO emission in NGC 4214. 

\cite{Braine2012} studied M 33 in the [\ion{C}{ii}] line with Herschel-HIFI at a spatial resolution of about 50 pc and also found that the [\ion{C}{ii}] lines are broader than the CO lines by about 50\% but narrower than \ion{H}{i}. Spatially resolved [\ion{C}{ii}] PACS maps \citep{Mookerjea2011} showed that there is little spatial correlation between [\ion{C}{ii}] and CO or \ion{H}{i}. 

The difference in [\ion{C}{ii}] and CO line profiles is not as striking as in M 33 or the star-forming regions in the SMC for the five regions of NGC 4214 presented in this paper. In regions I-1 and I-2, we find that the CO is broader by $\sim$ 20\%. Instead, in regions I-3, II, and III, the [\ion{C}{ii}] line is broader by $\sim 60$\%.
Although comparing the spectra does not directly show massive differences in the CO and [\ion{C}{ii}] spectra, our decomposition showed that the [\ion{C}{ii}] emission is associated not only with CO, but also with \ion{H}{i}.
Studying dwarf galaxies in different tracers with Herschel, \cite{Cigan2016} also found that \ion{H}{i} emission peaks correlate with [\ion{C}{ii}] peaks.  
We found that in region I the dominance (lowest value of (72$\pm$13)\% found in region I-1) of molecular mass is traced by [\ion{C}{ii}] rather than CO, implying extended CO-dark molecular gas. However, we note that the quoted values strongly depend on the assumed ISM conditions. Furthermore, we do not resolve  the gas clouds spatially as the SOFIA/GREAT beam of 14\arcsec corresponds to a spatial scale of about 200 pc.

Complementary to studies of low-metallicity environment, the origin of [\ion{C}{ii}] emission was also studied in the Milky Way. The Galactic Observations of Terahertz C$^{+}$ (GOTC$^{+}$; \citealt{GOTC+}) comprises Herschel-HIFI [\ion{C}{ii}] observations of long lines of sight through the Galactic plane. It was found that the [\ion{C}{ii}] emission correlates with the star formation rate on galactic scales and that its emission can be divided into contributions from different ISM phases. They attribute 30\% of the luminosity to dense PDRs, 25\% to cold \ion{H}{i}, 25\% to CO-dark H$_2$, and 20\% to ionized gas \citep{GOTCCII}. \cite{Velusamy2014} combined GOTC+ [\ion{C}{ii}] data with ancillary \ion{H}{i} and CO data to study the origin of Galactic [\ion{C}{ii}] emission. They found a widespread [\ion{C}{ii}] emission. About half of it is associated with diffuse molecular clouds faint in CO, but several lines of sight show features in either CO or \ion{H}{i} without corresponding emission in [\ion{C}{ii}].

The star-forming region M 17 SW was studied by \cite{Perez-Beaupuits2015} using spectrally resolved SOFIA/GREAT [\ion{C}{ii}] observations and by comparing them to [\ion{C}{i}] and low-$J$ CO lines. They analyzed their [\ion{C}{ii}] velocity channel maps with a different decomposition method and conclude that $\sim 65$\% of the mass found with [\ion{C}{ii}] is not associated with star-forming material. The decomposition of the spectra implies that about 36\%, 17\%, and 47\% of the [\ion{C}{ii}] emission originates from \ion{H}{ii}, \ion{H}{i}, and H$_2$ regimes, respectively.
Our study does not provide a comparison with a tracer of ionized gas, but previous studies showed that its contribution to the [\ion{C}{ii}] emission is low. Although there is large scatter, we find that we can associate about 46\% of the [\ion{C}{ii}] emission on average with the \ion{H}{i} emission.

\section{Summary}
We presented [\ion{C}{ii}] data observed with the GREAT instrument on board the SOFIA telescope in five regions of the low-metallicity irregular dwarf galaxy NGC 4214. The observations cover three separate regions in NGC 4214 in different evolutionary states. Star formation takes place in regions I and II whereas region III is considered less evolved with no star formation visible. We analyzed the velocity-resolved spectra by decomposing the [\ion{C}{ii}] emission spectra using ancillary \ion{H}{i} and CO(2 $\rightarrow$ 1) data. We estimated gas masses associated with different ISM phases. We summarize our results as follows:

\begin{itemize}
\item The \ion{H}{i} line profiles show almost the same widths of about 30 $\mathrm{km \, s}^{-1}$ in each of the five pointings and thus are broader than the [\ion{C}{ii}] lines by about 50 \% and broader than the CO lines by 20\% in regions I-1 and I-2. In regions I-3 and II, the \ion{H}{i} lines are more than twice as broad. All spectra peak at $\sim$ 300 $\mathrm{km \, s}^{-1}$, except for region III where the peaks are at $\sim$ 280 $\mathrm{km \, s}^{-1}$. The \ion{H}{i} peak positions are shifted to slightly lower velocities in three regions compared to the [\ion{C}{ii}]. In addition, the [\ion{C}{ii}] spectrum in this region is very noisy and thus difficult to decompose.

\item Using Gaussian curves to fit the [\ion{C}{ii}], CO, and \ion{H}{i} spectra we associate the majority ($> 70$ \%) of the emission with the CO emission in regions I-1 and I-2. In region I-3, about 65\% of the emission is attributed to the CO and in region II the fractions are split equally. On average, we can assign 54\% of the [\ion{C}{ii}] emission to the CO emission. 

\item The gas mass of atomic hydrogen is about 10$^6$ M$_\sun$ in each of the five pointings, the molecular gas mass traced by CO is $\sim$ 2 $\times$ 10$^5$ M$_\sun$. This is only about 25\% of the molecular mass seen in [\ion{C}{ii}]. The majority of molecular mass originates from CO-dark gas and between 32 and 87\% of the total [\ion{C}{ii}] intensity originates from this material. These results agree with most other studies in nearby dwarf galaxies. We find the highest fraction of CO-dark molecular gas mass of (89$\pm$10)\% in region I-2 and it coincides with the position of a super star cluster. In region I, the most evolved star-forming region, we find an average CO-dark gas fraction of 84\% while we find less CO-dark molecular gas (63\%) in region II. This region is younger, more metal-rich and more compact. The fraction of CO-dark molecular gas is therefore sensitive to the evolutionary stage of the regions.

\item Only about 9\% of the total [\ion{C}{ii}] emission can be attributed to the CNM in agreement with the findings from Local Group galaxies. There are indications that the \ion{H}{i}-associated [\ion{C}{ii}] also emits from denser phases. Similar to studies of the SMC and LMC, we find prominent wings in the [\ion{C}{ii}] profiles that cannot be fitted with CO spectra. The origin of these wings remains poorly constrained.
\end{itemize}

\begin{acknowledgements}
We would like to thank F. Walter for providing us with the OVRO CO(1-0) data, and also Bill Reach for his help in finishing this paper. We also thank the referee for constructive comments that helped improve the quality of the manuscript.
KF, DC, and FB acknowledge support from DFG grant BI 1546/1-1. This study is based on observations made with the NASA/DLR Stratospheric Observatory for Infrared Astronomy (SOFIA). SOFIA is jointly operated by the Universities Space Research Association, Inc. (USRA), under NASA contract NAS2-97001, and the Deutsches SOFIA Institut (DSI) under DLR contract 50 OK 0901 to the University of Stuttgart.

\end{acknowledgements}

\bibliographystyle{aa}
\bibliography{references}

\begin{appendix}
\section{Decomposition: Channel method}
\label{sec:appendix}

\subsection{Description of the method}
In conjunction with the decomposition method described in Section \ref{sec:Decomp} that uses Gaussians to fit the spectra, we also used the CO and \ion{H}{i} spectra directly. For each region, we want to describe our [\ion{C}{ii}] spectra $S$ by a linear combination of CO and \ion{H}{i}: 
\begin{equation}
S_{\mathrm{syn},\,[\ion{C}{ii}]} = a \, S_{\ion{H}{i}} + b \, S_{\mathrm{CO}}.
\label{eq:lincomb}
\end{equation}
This method shows some similarities with the approach that was used by \cite{Perez-Beaupuits2015} to study the origin of [\ion{C}{ii}] in M17 SW. 

To compute the synthetic spectra, we first normalize the CO and \ion{H}{i} spectra by their maximum and then choose the coefficients as percentages so that the CO and \ion{H}{i} spectra add up to 100\%. A percentage $X$ means that the whole \ion{H}{i} spectrum is multiplied by this number and then added to the CO spectrum multiplied correspondingly with 100-$X$\%.
In an ideal case, this combined synthetic spectrum would peak at the same flux density as the [\ion{C}{ii}] spectra. However, the best decomposition result might be achieved, if the synthetic spectra peaks at a slightly different flux density. Therefore, we introduce the scaling parameter $k$, where $k = 100\%$ corresponds to a scaling to the height of the [\ion{C}{ii}] peak. Equation \eqref{eq:lincomb} now reads
\begin{equation}
S_{\mathrm{syn},\,\mathrm{[\ion{C}{ii}]}} = k \, \mathrm{max}(S_{[\ion{C}{ii}]})\,\left( X\, \frac{S_{\mathrm{\ion{H}{i}}}}{\mathrm{max}(S_{\mathrm{\ion{H}{i}}})} + (100 - X) \, \frac{S_{\mathrm{CO}}}{\mathrm{max}(S_{\mathrm{CO}})} \right)
\label{eq:channeldecomp}
\end{equation} 

To quantify the quality of a decomposition with \ion{H}{i} percentage $X$ and scaling factor $k$, we compute the residual by subtracting the synthetic from the original [\ion{C}{ii}] spectrum. Tthe standard deviation of the residual is then used to measure the quality of this decomposition.

\begin{figure}
\centering
\includegraphics[width=0.24\textwidth]{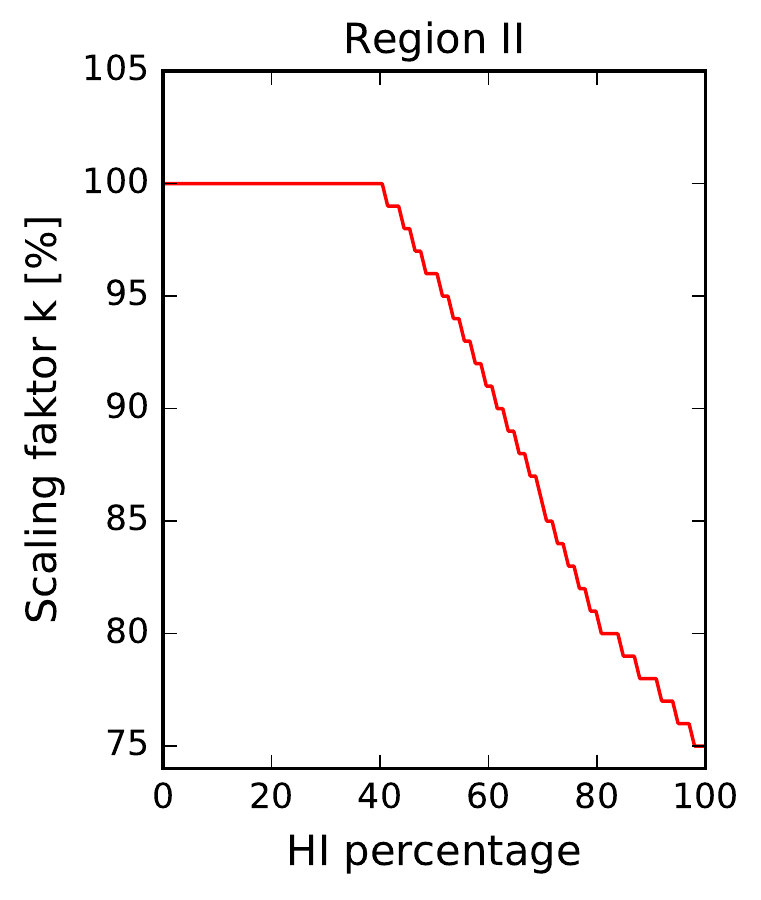}
\includegraphics[width=0.24\textwidth]{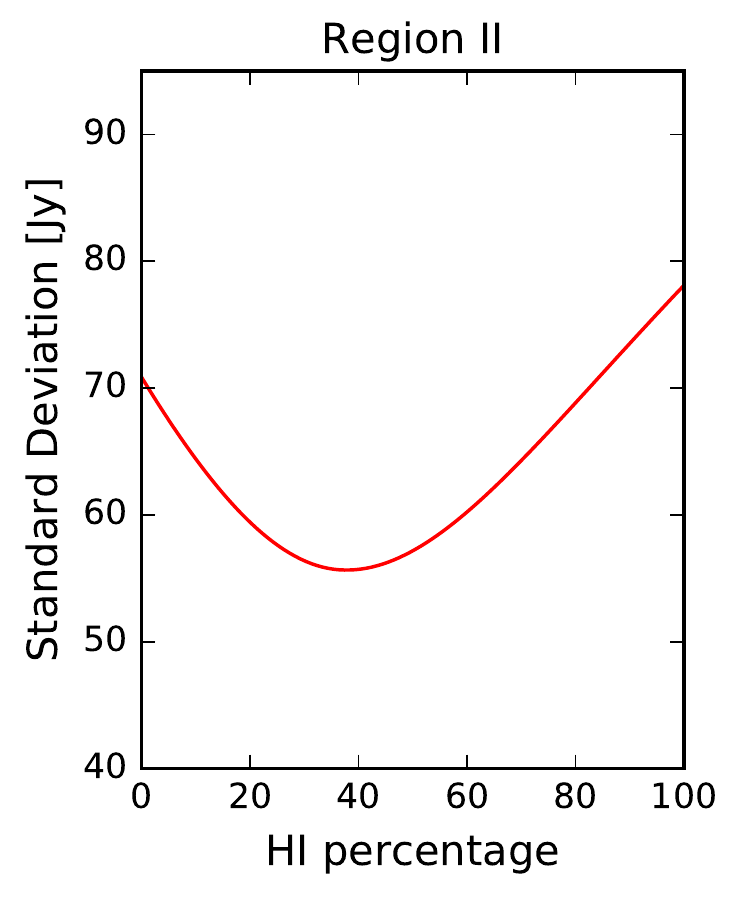} 
\caption{First panel: Scale factor $k$ that provides the lowest standard deviation of the residual versus the \ion{H}{i} percentage $X$ in region II. Second panel: Standard deviation versus \ion{H}{i} percentage. The minimum obtained for an \ion{H}{i} percentage of 37\% and a scaling factor of 100\%.}
\label{fig:thresholds}
\end{figure}

This procedure is repeated for all possible values of $X$ with integer percentages and scale factors ranging from 75\% to 100\% for each percentage.
For our best decomposition we want a low standard deviation of the residual and a high scaling factor $k$. To achieve the latter, we do not consider values of $k$ below 75\% since we want to reproduce the peak. We then vary the percentages $X$ and scaling factors $k$ and take the ($X$,$k$) pair with minimum residual (see Figure \ref{fig:thresholds}). In this example for region II, $X$\,=\,40\% and $k$\,=100\,.

However, this approach can result in residuals with broad negative features that would correspond to emission not seen in the [\ion{C}{ii}] spectra. These features are created mostly for high \ion{H}{i} percentages as the \ion{H}{i} spectra are broader than the [\ion{C}{ii}] spectra.

We filter out decompositions with these negative features by rejecting all  parameter pairs ($X,k$) that result in residuals with more than three consecutive velocity channels below a certain threshold.
For each pointing, this threshold is determined by the highest noise level of the [\ion{C}{ii}], CO(2\,$\rightarrow$\,1), and \ion{H}{i} spectra after scaling. While the RMS of the combined spectra depends on the CO and \ion{H}{i} spectra, the noise in the [\ion{C}{ii}] also sets a limitation for finding the best decomposition. We find that the best results are obtained when the threshold is set to 1$\sigma$ RMS of the noise. 

Using this approach, we can find parameter sets of $X$ and $k$ that give a 
decomposition for our five [\ion{C}{ii}] spectra. However, sometimes it is necessary to adjust the parameters manually, typically within 10\%, in order to get the best-matching synthetic spectrum. This range reflects the uncertainty of our channel decomposition method.

\subsection{Results}
\label{subsubsec:channelresults}
Table \ref{tab:percentages} reports the best results of the channel method decomposition. We note that the uncertainty of the \ion{H}{i} percentages are around 10 \%. The CO(2\,$\rightarrow$\,1) are 100\% minus the \ion{H}{i} percentages. We also note that the CO percentage factor is higher in all pointings. The [\ion{C}{ii}] line profile is more closely associated with the CO(2\,$\rightarrow$\,1) emission than with the broader \ion{H}{i} spectra, as was also seen in the Gaussian fit method. Table \ref{tab:percentages} also lists the [\ion{C}{ii}] intensity fractions associated with CO and \ion{H}{i}. These values were calculated by direct integration of the [\ion{C}{ii}] and the scaled CO and \ion{H}{i} spectra. 

\begin{figure*}
\includegraphics[width=1.\textwidth]{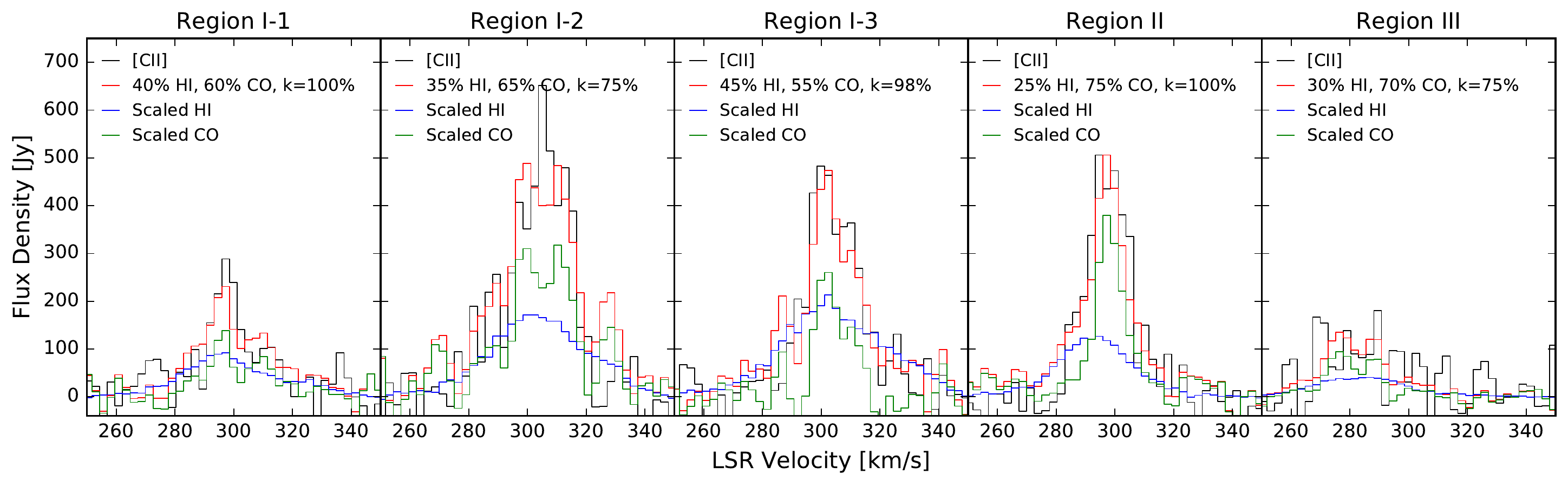} 
\includegraphics[width=1.\textwidth]{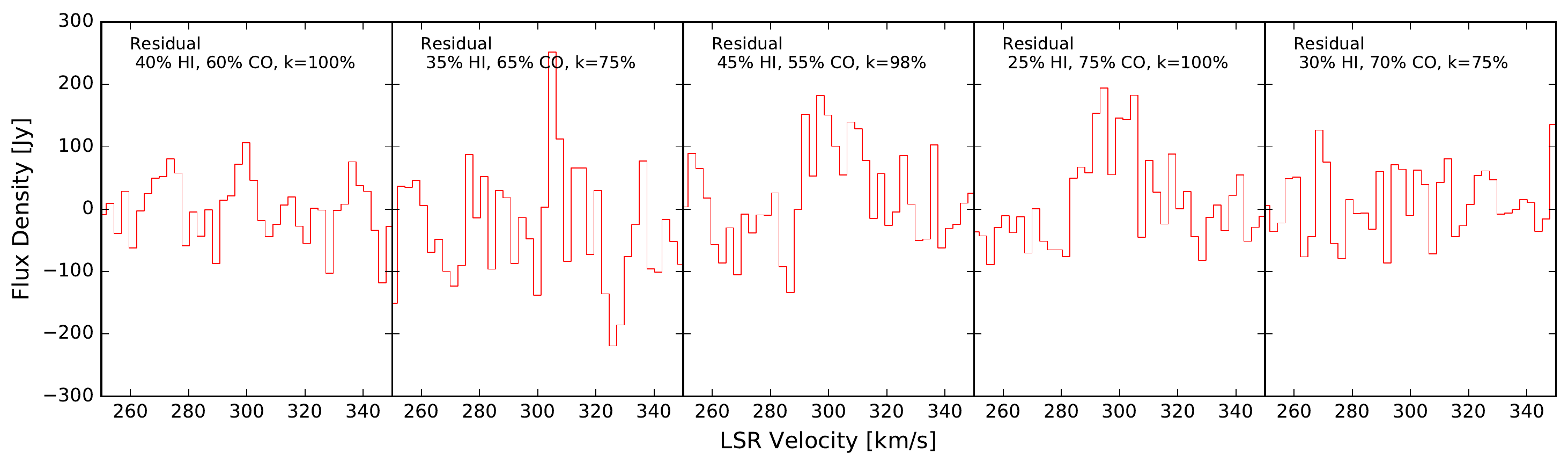} 
\caption{Channel method decomposition. Upper row: Original [\ion{C}{ii}] spectra in black and the best fitting synthetic spectra in red with the weightings given in Table \ref{tab:percentages}. The accordingly scaled \ion{H}{i} spectra are shown in blue, CO(2\,$\rightarrow$\,1) in green. Bottom row: Residuals computed from the weighted spectra by subtraction from [\ion{C}{ii}] data.}
\label{fig:result}
\end{figure*}

Figure \ref{fig:result} presents the final decomposition of our [\ion{C}{ii}] spectra with the corresponding residuals. The decomposition works best for region I-1 and region II where the weighted spectra of \ion{H}{i} and CO reproduce the shape of the [\ion{C}{ii}] well. Region III shows a flat residual, but here the best fitting decomposition does not match well owing to the broad, uneven line shape. In this region, it is questionable whether the [\ion{C}{ii}] emission is strong enough to be decomposed in a useful manner. The decomposition of [\ion{C}{ii}] in region I-3 looks reasonable, although the residual shows a significant dip at $\sim 290$ km/s. We note that it was not possible to reduce this dip any further.
The decomposition is worst in region I-2. The weighted spectrum shows side features and has a broad top with two peaks. 

Overall, the weighted and scaled spectra of \ion{H}{i} and CO(2\,$\rightarrow$\,1) can account for most [\ion{C}{ii}] emission at least reasonably well. We find no strong indication for components in the observed regions that are not visible in either \ion{H}{i} or CO, but are visible in [\ion{C}{ii}].

\begin{table*}
\centering
\begin{threeparttable}
\caption{Results from the channel method decomposition}.\vspace{-10pt}
\begin{tabular}{c c c c c c} \hline \hline
Region & \ion{H}{i} & CO(2\,$\rightarrow$\,1) & $k$ & $I_{[\ion{C}{ii}]}(\ion{H}{i})$/$I_{[\ion{C}{ii}]}$ & $I_{[\ion{C}{ii}]}\mathrm{(CO)}$/$I_{[\ion{C}{ii}]}$\\ \hline
I-1 & 40\% & 60\% & 80\% & 55.0\% & 44.1\%\\
I-2 & 35\% & 65\% & 75\% & 47.8\% & 61.9\% \\
I-3 & 45\% & 55\% & 98\% & 70.6\% & 28.7\%\\
II & 25\% & 75\% & 100\% & 40.4\% & 61.9\%\\
III & 30\% & 70\% & 75\% & 37.5\% & 39.1\%\\ 
\hline
\end{tabular}
\begin{tablenotes}
\item\small Notes. Intensity fractions $I_{[\ion{C}{ii}]}(\ion{H}{i})$/$I_{[\ion{C}{ii}]}$ and $I_{[\ion{C}{ii}]}(\mathrm{CO})$/$I_{[\ion{C}{ii}]}$ using the scaled \ion{H}{i} and CO spectra.
\end{tablenotes}
\label{tab:percentages}
\end{threeparttable}
\end{table*}

When comparing the results to those of the Gaussian method (see Table \ref{tab:fitresults}), we find no significant differences. However, we prefer the Gaussian fitting method because it enables us to fit several components in the \ion{H}{i} spectra. In addition, the channel method requires manual tuning to find the best parameter set for the synthetic [\ion{C}{ii}] spectrum. 


\end{appendix}

\end{document}